\documentclass{article}
\usepackage{enumitem}
\usepackage{hyperref}
\usepackage{graphicx}
\usepackage{xcolor}

\title{%
    Critical Canvas \\
  \large How to regain information autonomy in the AI era}
\author{Dong Chen}
\date{\today}

\begin{document}

\maketitle

\tableofcontents
\newpage

\section{Background and Motivation}

The rise of AI has fundamentally transformed how we consume and process information. While AI-powered systems have made information more accessible than ever, they have also created unprecedented challenges to our information autonomy. Two critical developments particularly threaten our ability to form independent thoughts and make informed decisions:

First, recommendation algorithms create personalized echo chambers, efficiently distributing content that reinforces existing beliefs while limiting exposure to diverse perspectives. Second, the emergence of generative AI has blurred the line between authentic and fabricated content, making it increasingly difficult to distinguish truth from fiction. The combination of these forces threatens not just individual autonomy but the very foundation of democratic discourse.

This challenge is particularly critical in the domain of technical AI governance, where stakeholders must navigate intricate technical specifications, safety frameworks, and implementation details. As AI systems become more complex, the ability to understand and evaluate technical documentation, research papers, and implementation guidelines becomes crucial for effective governance.

The Critical Canvas addresses these challenges by providing a novel information exploration and visualization platform. It enables users to navigate topics ranging from high-level overviews to nuanced specifications across multiple dimensions, evaluate source credibility, and build comprehensive understanding through structured yet flexible exploration pathways. Think of it as a digital canvas where users can zoom in to examine specific details, zoom out to grasp broader contexts, and track the credibility of information sources - all while maintaining their agency in the exploration process.

\section{Product Overview}

The Critical Canvas is designed to address these challenges by restoring balance between algorithmic efficiency and human agency. Through its structured yet flexible approach to information exploration, it transforms overwhelming information landscapes into actionable insights.

\subsection{Core Functionality}
The system achieves this through three primary mechanisms:

\begin{itemize}[itemsep=0.3cm]
    \item \textbf{Multi-dimensional Navigation}: Users can explore complex topics across temporal, geographical, and logical dimensions, enabling both broad understanding and deep technical insight.
    
    \item \textbf{Source Credibility Tracking}: A sophisticated system evaluates and tracks source reliability, helping users navigate technical documentation and research with confidence.
    
    \item \textbf{Knowledge Integration}: Dynamic knowledge entries connect related concepts, specifications, and implementations, creating a comprehensive view of technical AI governance topics.
\end{itemize}

These capabilities allow users to:
\begin{itemize}[itemsep=0.3cm]
    \item Gain high-level overviews while accessing detailed technical specifications
    \item Navigate seamlessly between related concepts and their implementations
    \item Verify the credibility of technical sources and research
    \item Save, share, and reuse their exploration pathways for collaborative governance efforts
\end{itemize}

\subsection{Target Users}

The tool caters to a primary user group of \textbf{Comprehensive Knowledge Seekers} – individuals who aim to gain a holistic understanding of complex stories or topics. These users value the flexibility to explore narratives across multiple dimensions and the transparency to evaluate sources and their contexts.

Within this broader category, the following subclasses illustrate specific user types:

\begin{itemize}[itemsep=0.3cm]
    \item \textbf{Curiosity-Driven Explorers}: Individuals seeking to deepen their understanding of complex topics in a structured yet intuitive manner.
    \item \textbf{Professionals and Researchers}: Policymakers, analysts, and academics who require rigorous exploration pathways, credible sources, and tools to share findings with teams or communities.
\end{itemize}

By addressing the needs of these groups, the tool ensures a seamless experience for users ranging from casual explorers to systematic analysts, enabling effective navigation through modern information landscapes.

\section{Key Components}
The following components form the foundational building blocks of the system, enabling the core features that users interact with:

\subsection{Knowledge Entry (KE)}
\subsubsection{Definition and Structure}
A Knowledge Entry serves as the fundamental unit of information organization in the Critical Canvas. Each KE functions as a dynamic container that captures knowledge about a specific concept, event, or topic across three primary dimensions: logical, temporal, and geographical. This multi-dimensional structure allows the system to represent knowledge in its natural form, accommodating both broad concepts and specific instances.

For example, consider the concept of AI Safety. As a KE, it contains logical aspects such as technical approaches and theoretical frameworks, temporal elements tracking its evolution from early research to current developments, and geographical variations in implementation and regulation across different regions. This structure ensures that users can access comprehensive information while maintaining clear contextual boundaries.

\subsubsection{KE Relationships and Dynamics}
Knowledge Entries form an interconnected network that reflects the natural relationships between concepts and ideas. These relationships manifest in several ways:

The primary relationship type is hierarchical containment, where more specific KEs nest within broader ones. For instance, while "AI Safety" represents a broad field, it contains more focused KEs like "AI Alignment" or "Robustness." These contained KEs inherit relevant context from their parent while maintaining their distinct focus and characteristics. This hierarchy can extend multiple levels deep, such as "AI Alignment Research in China" being contained within "AI Alignment," which itself sits within "AI Safety."

KEs can also emerge through dimensional constraints applied to existing entries. When users focus on specific time periods or geographical regions, the system generates new KEs that preserve the relevant subset of information while maintaining connections to the broader context. For example, "AI Safety in the EU" represents a geographical constraint on the broader AI Safety KE, while "AI Safety in the EU post-2020" adds a temporal constraint to further refine the scope.

Beyond hierarchical relationships, KEs maintain cross-references that capture important connections without implying containment. These bidirectional links allow users to explore related concepts naturally. For instance, the "AI Governance" KE might reference both "AI Safety" and "Policy Making" KEs, reflecting the interdisciplinary nature of the field.

This rich network of relationships enables the system to maintain consistency during updates while allowing for dynamic growth as users discover new connections and relationships through their explorations.

\subsection{Source Credibility System}
\begin{itemize}[itemsep=0.3cm]
    \item \textbf{Content Evaluation Framework}:
    \begin{itemize}
        \item \textit{Evidence Assessment}:
        \begin{itemize}
            \item Consensus indicators across reliable sources
            \item Verification status and fact-checking results
            \item Cross-reference metrics with independent confirmations
            \item Update history and corrections
            \item Supporting data and methodology transparency
        \end{itemize}
        
        \item \textit{Narrative Analysis}:
        \begin{itemize}
            \item Balance of fact-based vs narrative-based content
            \item Sensationalism vs objectivity metrics
            \item Emotional language assessment
            \item Contextual completeness
            \item Bias indicators in framing and presentation
        \end{itemize}
    \end{itemize}

    \item \textbf{Source Phase Space}: A dynamic system for evaluating content creators and institutions:
    \begin{itemize}
        \item Historical track record in the same domain
        \item Expertise areas and credentials
        \item Publication patterns and methodologies
        \item Institutional affiliations and potential biases
        \item Response to corrections and criticism
    \end{itemize}

    \item \textbf{Integration}: The system combines Content Evaluation and Source Phase Space to:
    \begin{itemize}
        \item Provide context for how reliable sources handle evidence and narrative
        \item Track patterns in source reliability across different types of content
        \item Build comprehensive credibility profiles over time
    \end{itemize}
\end{itemize}

\subsection{Query System}
\subsubsection{User Interaction}
The tool’s query system serves as the entry point for user interaction. Users can input broad or specific questions, which the system processes to identify the most relevant KE and dimensions to explore.

\subsubsection{Query Processing}
The system interprets the query using a structured taxonomy, ensuring accurate identification of the core object (KE) and related attributes.

\subsubsection{Focus Mechanism (Taxonomy)}
Taxonomy serves as the underlying focus mechanism, guiding the system to:
\begin{itemize}[itemsep=0.3cm]
    \item Break down user queries into logical components.
    \item Retrieve and organize the most relevant information.
    \item Present a curated and structured response to the user.
\end{itemize}

\subsection{Navigational Pathways}
The Navigational Pathways component serves as the system's memory and learning mechanism, capturing and structuring users' exploration journeys. Each pathway is recorded as a rich data structure that preserves not only the sequence of interactions but also the context and relationships discovered during exploration. The system maintains these pathways as directed graphs, where nodes represent interaction points - such as queries, content views, and source evaluations - while edges capture the logical flow and relationships between these interactions.

The component implements sophisticated versioning and inheritance mechanisms, allowing pathways to evolve and branch while maintaining their historical context. When users modify or extend existing pathways, the system preserves the original exploration pattern while accommodating new insights and connections. This architectural approach enables the system to learn from collective user behavior, identifying common exploration patterns and potential knowledge gaps that inform future pathway recommendations.

\subsection{Exploration Pathways and History}
The Critical Canvas transforms every exploration journey into a reusable and shareable asset through its pathway feature. As users navigate through topics, their interactions naturally create a structured record of their exploration journey. This pathway captures not just the sequence of steps taken, but also the insights gained, sources evaluated, and connections discovered along the way.

Users can revisit their pathways at any time, picking up exactly where they left off or branching into new directions of inquiry. For instance, a researcher exploring AI safety might create a pathway focusing on technical aspects, then later return to explore policy implications using the same foundational understanding. These pathways become particularly valuable when shared with colleagues or the broader community, as others can follow, extend, or adapt the exploration pattern while maintaining attribution to the original insights.

The system also learns from these collective exploration patterns, suggesting potentially relevant branches or alternative perspectives that others have found valuable. This creates a dynamic knowledge ecosystem where individual explorations contribute to and benefit from the community's collective understanding. Whether used for personal research, team collaboration, or public education, pathways serve as both a record of discovery and a launch point for deeper understanding.

\section{Core Features}
The Critical Canvas transforms its technical infrastructure into intuitive user experiences through three core features. Each feature directly leverages one or more key components to help users navigate, understand, and share complex knowledge effectively.

\subsection{Multi-Dimensional Navigation and Visualization}
At the heart of the user experience lies a sophisticated navigation system that makes complex topics accessible through three complementary dimensions. The logical dimension allows users to naturally decompose topics into manageable concepts. When exploring AI safety, for instance, users can seamlessly navigate from broad concepts to specific aspects like value alignment or robustness, with each concept presented through interactive concept cards and relationship visualizations that clearly show connections and dependencies.

The temporal dimension brings historical context to life through dynamic timelines. Users can trace the evolution of ideas, marking significant milestones and understanding how concepts have developed over time. Interactive elements within these timelines allow users to explore key events in detail, understanding not just what happened but also its implications and connections to other developments. For example, in the AI safety timeline, users can see how the release of ChatGPT influenced subsequent policy discussions and technical research directions.

The geographical dimension contextualizes knowledge across different regions and cultures through interactive maps and comparative visualizations. Users can explore how approaches and implementations vary globally, understanding regional priorities and challenges. In the context of AI safety, this might reveal how different regions approach regulation, with the United States favoring corporate-led initiatives while the European Union pursues a more regulation-first strategy.

\subsection{Personalized User Experience}
The system adapts to each user's exploration patterns and learning needs, creating a truly personalized journey through complex topics. As users navigate through different dimensions, the system learns from their interactions to provide relevant recommendations and insights. These suggestions help users discover related concepts they might find interesting or important, while maintaining a clear connection to their original query or area of interest.

The personalization extends beyond content recommendations to include customizable visualization preferences and exploration paths. Users can adjust how information is presented, save important insights for later reference, and create custom views that align with their specific research or learning objectives. This flexibility ensures that whether someone is conducting academic research or seeking general knowledge, they can interact with information in a way that suits their needs.

\subsection{Sharing and Collaboration}
Knowledge exploration becomes a collaborative endeavor through robust sharing and teamwork features. Users can share their entire exploration pathways, complete with annotations, insights, and custom visualizations, enabling others to benefit from their discoveries and perspective. These shared pathways serve as more than just static records; they become interactive templates that others can build upon and adapt to their own needs.

The collaboration features support both asynchronous and real-time interaction. Teams can work together in shared workspaces, annotating findings, discussing implications, and building collective understanding. This collaborative environment is particularly valuable for research teams, policy makers, and other groups working to understand complex topics from multiple angles. The system maintains the provenance of insights and contributions, allowing teams to track how their collective understanding evolves over time.

\section{User Workflow}

\subsection{Initial Query}
Users begin by entering a query or selecting a predefined question from a curated list of examples. The query is processed to identify the relevant Knowledge Entry (KE) and dimensions (logical, temporal, and geographical).

\subsection{Navigation and Exploration}
Once the KE is identified, users explore it through interactive zooms:
\begin{itemize}[itemsep=0.3cm]
    \item \textbf{Logical Zoom}: Breaks the topic into sub-concepts and related fields.
    \item \textbf{Temporal Zoom}: Highlights historical milestones, current trends, and future projections.
    \item \textbf{Geographical Zoom}: Displays global or regional insights through interactive maps.
\end{itemize}

\subsection{KE Generation and Refinement}
Dynamic KEs are created or refined based on user queries and interactions. For new areas, the system generates initial entries using its database and links to credible sources.

\subsection{Session Archiving}
The exploration pathway—including queries, zoom interactions, and source checks—is saved as a session. Users can revisit and refine these sessions at any time.

\subsection{Reusability and Sharing}
Saved sessions can be reused for similar queries or shared with others. These pathways also contribute to enhancing the system’s knowledge base, making it easier to handle future queries.

\section{Use Case: Alex's Story}

\subsection*{Setting the Scene}

Alex, a journalist deeply passionate about environmental policy, decides to investigate the global implications of AI safety. The subject is vast and often confusing, with numerous stakeholders, technical jargon, and conflicting viewpoints. Alex feels overwhelmed trying to make sense of scattered information online and wants a holistic tool to navigate the topic.

\subsection*{Starting the Journey}

Alex logs into The Critical Canvas and enters the query:

\textit{``What are the global risks and governance challenges associated with AI safety in the 21st century?"}

The tool processes the query and identifies \textbf{AI Safety} as the primary \textbf{Knowledge Entry (KE)}. It highlights dimensions such as logical sub-concepts, historical timelines since the 21st century, and geographical perspectives. Alex starts exploring.

\subsection*{Exploring Logical Zoom}

Alex clicks on the Logical Zoom, curious about the foundational aspects of AI safety. The tool breaks the topic into manageable sub-concepts like \textit{Value Alignment}, \textit{Robustness}, and \textit{Ethics and Governance}. Alex selects \textit{Ethics and Governance}, the tool dynamically navigates to the corresponding KE, and gives Alex an overview of how ethics and governance influence AI safety.

\subsection*{Diving into the Temporal Zoom}

Alex wants to go back to the layer of AI Safety, thus zooms out from \textit{ethics and governance}. Then, using the Temporal Zoom, Alex views a timeline of AI safety milestones, including: 

\begin{itemize}[itemsep=0.3cm]
    \item \textbf{2013}: Nick Bostrom's ``Superintelligence'' published
    \item \textbf{2015}: Open Letter on AI Safety signed by prominent researchers
    \item \textbf{2015}: OpenAI founded with explicit focus on beneficial AI
    \item \textbf{2016}: AlphaGo beats Lee Sedol
    \item \textbf{2017}: Asilomar AI Principles established
    \item \textbf{2018}: Google's AI Principles published
    \item \textbf{2019}: GPT-2 release delayed due to misuse concerns
    \item \textbf{2022}: ChatGPT release sparks widespread AI safety discussions
    \item \textbf{2023}: ``AI Pause Letter'' signed by tech leaders
    \item \textbf{2023}: Anthropic's Constitutional AI approach
    \item \textbf{2023}: AI Safety Summit at Bletchley Park
\end{itemize}

and ongoing efforts to collaborate internationally, such as at conferences and summits. Each milestone provides clickable insights into associated implications and shortcomings.

\subsection*{Navigating Geographical Perspectives}

Alex wonders how countries differ in their approach. The Geographical Zoom displays an interactive map highlighting:

\begin{itemize}[itemsep=0.3cm]
    \item \textbf{The U.S.}: Corporate-led initiatives.
    \item \textbf{Europe}: Regulation-first strategies.
    \item \textbf{China}: State-aligned frameworks.
\end{itemize}

This exploration reveals regional challenges, ethical debates, and geopolitical tensions in AI governance.

\subsection*{Evaluating Credibility with the Source Phase Space}

As Alex explores, the tool highlights sources contributing to the KE. Using the \textbf{Source Phase Space}, Alex examines institutions (Future of Life Institute, DeepMind), individuals (Stuart Russell, Dario Amodei), and publications (peer-reviewed articles). Credibility indicators, such as sensationalism and fact-based metrics, help Alex opt out of non-credible sources. Brief notes document why certain sources were excluded.

\subsection*{Saving and Sharing Pathways}

After hours of exploring, Alex saves the entire session as a \textbf{Navigational Pathway}. The pathway includes the query, key insights, source evaluations, and notes. Alex shares a summary report with visualizations, timelines, and credibility data with friends and colleagues. Their feedback enhances Alex’s understanding and refines the article further.

\subsection*{Reflecting on the Experience}

The Critical Canvas transforms Alex’s overwhelming topic into actionable insights. The structured approach empowers Alex to:

\begin{itemize}[itemsep=0.3cm]
    \item Understand AI safety comprehensively.
    \item Evaluate sources confidently.
    \item Collaborate effectively with colleagues.
\end{itemize}

Alex now considers The Critical Canvas indispensable for investigative journalism, enabling deep dives into complex topics with unparalleled clarity. For AI safety, he plans to spend another session on his remaining questions after the discussion with his colleagues.

\section{Conclusion}

The Critical Canvas represents a novel approach to information consumption in the AI era, built upon three fundamental principles:

Creative information visualization transforms complex technical knowledge into an intuitive, multi-dimensional space. By representing information across logical, temporal, and geographical dimensions, the system enables users to discover patterns and relationships that might otherwise remain hidden in traditional linear presentations.

This visualization framework supports human-centric navigation, allowing users to explore information at their own pace and according to their own interests. Unlike algorithm-driven recommendations that push users down predetermined paths, the Critical Canvas empowers users to chart their own course through information landscapes, from high-level overviews to nuanced specifications.

Most importantly, the system helps users regain autonomy over their information consumption by making the entire process transparent and examinable. Through its Source Credibility System and structured exploration pathways, users can understand not just what they're learning, but why they should trust it and how it connects to their existing knowledge.

Together, these principles create a tool particularly suited for technical AI governance, where the ability to navigate complex information landscapes while maintaining critical perspective is crucial. The Critical Canvas demonstrates that by combining creative visualization, human-centric design, and transparency, we can build tools that enhance rather than diminish human agency in the age of AI.

\end{document}